\documentclass[12pt,preprint]{aastex}

\newcommand\jetp{{Sov. Phys. JETP}}

\newcommand\pop{{Phys.~Plasmas}}
\newcommand\rpp{{Rep.~Progr.~Phys.}}
\newcommand\ass{{Astroph. Space Sci.}}

\newcommand\etal{{\it et al.,\ }} % macros exists
\newcommand\eg{{e.g.,\ }}
\newcommand\ie{{ i.e.,\ }}
\newcommand\jpg{{Journ. Phys. G}}
\newcommand\const{{\rm const\ }}

\shortauthors{Malkov, Diamond \protect\& Jones}
\shorttitle{TeV protons in SNRs}
\begin{document}

\title{On the possible reason for non-detection of TeV-protons in SNRs}

\author{M. A. Malkov and P. H. Diamond}

\affil{University of California at San Diego, La Jolla, CA 92093-0319}
\email{mmalkov@ucsd.edu}

\and{}

\author{T. W. Jones}

\affil{University of Minnesota, Minneapolis, MN 55455}

\begin{abstract}
The theory of shock acceleration predicts the maximum particle energy
to be limited only by the acceleration time and the size (geometry)
of the shock. This led to optimistic estimates for the galactic cosmic
ray energy achievable in the SNR shocks. The estimates imply that
the accelerated particles, while making \emph{no strong impact on
the shock structure} (test particle approach) are nevertheless scattered
by \emph{strong self-generated} Alfven waves (turbulent boost) needed
to accelerate them quickly. We demonstrate that these two assumptions
are in conflict when applied to SNRs of the age required for cosmic
ray acceleration to the {}``knee{}'' energy.

We study the \emph{combined} effect of acceleration nonlinearity (shock
modification by accelerated particles) and wave generation on the
acceleration process. We show that the refraction of self-generated
waves resulting from the deceleration of the plasma flow by the pressure
of energetic particles causes enhanced losses of these particles.
This effect slows down the acceleration and changes the shape of particle
spectrum near the cut-off. The implications for observations of TeV
emission from SNR remnants are also discussed.
\end{abstract}


\section{Introduction}

The first-order Fermi or diffusive shock acceleration (DSA) has been
long considered as responsible for the production of galactic cosmic
rays (CRs) in supernova remnants (SNRs), as well as for the radio,
\( x \)- and \( \gamma  \)-ray emission from these and other shock
related objects. The most crucial characteristic of this process that
is usually examined in terms of its capability to explain a given
observation, is the rate at which it operates. Indeed, what is often
expected from the theory or even inferred from the observations is
an extended particle energy spectrum, frequently a power-law, but
more rapidly decaying at the highest energies observed. Often, this
decay is referred to as an energy or momentum cut-off and is usually
associated with the finite acceleration time or with losses if their
rate exceeds the acceleration rate. As long as the losses are unimportant,
the cut-off \( p_{max}(t) \) advances with time according to the
following equation 

\begin{equation}
\label{p:max}
\frac{dp_{max}}{dt}=\frac{p_{max}}{t_{acc}}
\end{equation}
 whereas in the presence of losses the acceleration rate \( p_{max}/t_{acc} \)
may be equated to the loss rate to yield a steady state value of \( p_{max} \).
The acceleration time scale is determined by (\eg \citealt{axf81})

\begin{equation}
\label{t:acc}
t_{acc}=\frac{3}{u_{1}-u_{2}}\int _{p_{min}}^{p_{max}}\left[ \frac{\kappa _{1}(p)}{u_{1}}+
\frac{\kappa _{2}(p)}{u_{2}}\right] \frac{dp}{p}
\end{equation}
 with \( u_{1} \) and \( u_{2} \) being the upstream and downstream
flow speeds in the shock frame and with \( \kappa _{1,2} \) being
the particle diffusivities in the respective media. One may recognize
in the last formula the sum of average residence times of a particle
spent upstream and downstream of the shock front before it completes
one acceleration cycle, integrated over the entire acceleration history
from \( p_{min} \) to \( p_{max} \). Given the flow speeds \( u_{1,2} \)
which, in many cases are known reasonably well, the most sensitive
quantity is the particle diffusivity \( \kappa  \). This, in turn,
is determined by the rate at which particles are pitch angle scattered
by the Alfven turbulence. If the latter was just a background turbulence
in the interstellar medium (ISM), the acceleration process would be
too slow to produce the galactic CRs in SNRs (\eg \citealt{lc83}). However it was realized
(\eg \citealt{bell78a,bla:ost}) that accelerated particles should create the
scattering environment by themselves generating Alfven waves on the
cyclotron resonance \( kp\mu /m=\omega _{ci} \), where \( k \) is
the magnitude of the wave vector (directed along the magnetic field),
\( p \), \( \mu  \), \( m \) and \( \omega _{ci} \) are the particle
momentum, the cosine of its pitch angle, mass and non-relativistic
(\( eB/mc \)) gyro-frequency. Note, that the diffusive character
of particle transport (and determination of \( \kappa  \)) has been
rigorously obtained within a quasi-linear theory, \ie it is subject
to constraints on the turbulence level.

The wave generation, however, proved to be very efficient (see \eg \citealt{vdm84}
and the next section). In particular, using, again, the quasi-linear
approximation, the normalized wave energy density \( \left( \delta B/B_{0}\right) ^{2} \)
may be related to the partial pressure \( P_{c} \) of CRs that resonantly
drive these waves through

\begin{equation}
\label{delB}
\left( \delta B/B_{0}\right) ^{2}\sim M_{A}P_{c}/\rho u^{2}
\end{equation}
 where \( M_{A} \) is the Alfven Mach number and \( \rho u^{2} \)
is the shock ram pressure. Since \( M_{A} \) is typically a large
parameter, \( \delta B/B_{0} \) may become larger than unity even
if the acceleration itself is relatively inefficient, \ie if \( P_{c}/\rho u^{2}\ll 1 \).
Strictly speaking this invalidates the quasi-linear approach as a
means for describing the generation of strong turbulence at shocks.
The commonly accepted way to circumvent this difficulty is to assume
that the turbulence saturates at \( \delta B/B_{0}\sim 1 \), which
means that the m.f.p. of pitch angle scattered particles is of the
order of their gyro-radius \( r_{g} \). Then, \( \kappa =\kappa _{B}\equiv c \)\( r_{g}(p)/3 \),
where the speed of light \( c \) is substituted instead of CR velocity
and \( \kappa _{B} \) stands for the Bohm diffusion coefficient.
This immediately sets the acceleration time scale (\ref{t:acc}) at
the level of particle gyro-period \( \left( eB/p\right) ^{-1} \)
times \( \left( c/u_{1}\right) ^{2} \). In principle, the turbulence
level \( \delta B/B_{0} \) significantly exceeding unity is possible
in local shock environments (see \eg numerical studies by \citealt{be95}
and \citealt{bell:luc}). As a consequence of that the diffusion coefficient
could be even smaller than \( \kappa _{B} \), and hence, the acceleration
rate would be faster than it is commonly believed to be. At the same
time, since usually Alfvenic type turbulence is considered, the respective
velocity perturbations must be super-Alfvenic and supersonic, which
raises questions about its ability to sustain itself in an extended
area without rapid dissipation that will decrease the \( \delta B/B_{0} \)
level.  Likewise, decreasing of turbulence level below the  Bohm limit, for example due
to the finite extent of the turbulence zone upstream, should slow down
the acceleration \citep{lc83}.

However the acceleration rate given by eq.(\ref{t:acc}) with \( \kappa =\kappa _{B} \)
was found to be fast enough to explain (at least marginally)
the acceleration of CRs in SNRs up to the {}``knee{}'' energy \( \sim 10^{15}eV \)
over their life time. Much further optimism has been caused by the
studies of \citet{dav} and \citet{naito}. They analyzed the prospects for
detection of super-TeV emission from nearby SNR that should be
produced by the decays of \( \pi ^{0} \) mesons born in collisions
of shock accelerated protons with the nuclei of interstellar gas.
The expected fluxes were shown to be detectable by the imaging \( \rm \check{C} \)erenkov
telescopes. Moreover, the EGRET \citep{esp} detected a lower energy
(\( \la GeV \)) emission coinciding with some galactic SNRs. The
spectra also seemed consistent with the DSA predictions. One may even
argue that the low energy EGRET data verified one of the most difficult
elements of the entire acceleration mechanism, the so called injection.
In essence, this is a selection process (not completely understood)
whereby a small number of thermal particles become subject to further
acceleration (see \citealt{gjk00,zank01} for the latest development
of the injection theory and \citealt{mdru01} for a review) and may be
then treated by standard means of the DSA theory that was designed
to describe particles with velocities much higher than the shock velocity.
Therefore, what seemed left for the theory was to continue the EGRET
spectrum (that sets the normalization constant, or injection rate)
with some standard DSA slope (nearly \( E^{-2} \) or somewhat steeper)
and to predict the \( \gamma  \)-ray flux in the TeV range where
it could be detected by the \( \rm \check{C} \)erenkov telescopes.

Unfortunately, despite the physical robustness of the arguments given
by  \cite{dav,naito}, no statistically significant signal
that could be attributed to any of the EGRET sources was detected.
The further complication is that some critical energy band between
GeV and TeV energies is currently uncovered by available instruments.
Therefore, based on these observational results it was suggested (\eg
\citealt{buck98}) that there is probably a spectral break or even cutoff
somewhere within this band. However the spectrum above GeV energies
remains an enigma. This will be resolved perhaps with the launch of
the GLAST mission and when the new generation of \( \rm \check{C} \)erenkov
telescopes with lower energy thresholds begin to operate. However,
the discovery of the 100 TeV emission from SNR1006 \citep{tanim98},
as well as some other remnants not seen by the EGRET at lower energies
(see, \eg \citealt{ahar01,allen01,kirkd01} for a complete discussion), although
almost universally identified with electrons diffusively accelerated
to similar energies,  is widely interpreted as  a strong support of  the mechanism itself.
The above suggests, however, that in reality it might be not as robust as is its simplified
test particle version with enhanced turbulence and particle scattering.

In this paper we attempt to understand what may happen to the spectrum
provided that the acceleration is indeed fast enough to access the
TeV energies over the life time of SNRs in question. Our starting
point is that the fast acceleration also means that the pressure of
accelerated particles becomes significant in an early stage of Supernova
evolution so that the shock structure is highly nonlinear. At the
first glance this should not slow down acceleration since, according
to eq.(\ref{delB}), this changes the turbulence \emph{level} thus
improving particle confinement near the shock front and thus making
acceleration faster (smaller \( \kappa  \)). However, the formation
of a long CR precursor (in which the upstream flow is gradually decelerated
by the pressure of CRs, \( P_{c} \)) influences the \emph{spectral
properties} of the turbulence by affecting the propagation and excitation
of the Alfven waves. This effect is twofold. First the waves are compressed
in the converging plasma flow upstream and are thus blue-shifted,
eliminating the long waves needed to keep exactly the highest energy
particles diffusively bound to the accelerator. Second, and as a result
of the first, at highest energies there remain fewer particles than
expected so that the level of resonant waves is smaller and hence
the acceleration rate is lower. We believe that these effects have
been largely overlooked before which may have substantially overestimated
the particle maximum energy in strongly nonlinear regimes.

\section{Basic Equations and Approximations}

We use the standard diffusion-convection equation for describing the
transport of high energy particles (CRs) near a CR modified shock.
First, we normalize the distribution function \( f(p) \) to \( p^{2}dp \):

\begin{equation}
\label{dc1}
\frac{\partial f}{\partial t}+U\frac{\partial f}{\partial x}-\frac{\partial }{\partial x}\kappa \frac{\partial f}{\partial x}=
\frac{1}{3}\frac{\partial U}{\partial x}p\frac{\partial f}{\partial p}
\end{equation}
 Here \( x \) is directed along the shock normal which for simplicity
is assumed to be the direction of the ambient magnetic field. The
two quantities that control the acceleration process are the flow
profile \( U(x) \) and the particle diffusivity \( \kappa (x,p) \).
The first one is coupled to the particle distribution \( f \) through
the equations of mass and momentum conservation

\begin{eqnarray}
\frac{\partial }{\partial t}\rho +\frac{\partial }{\partial x}\rho U=0 &  & \label{mas:c} \\
\frac{\partial }{\partial t}\rho U+\frac{\partial }{\partial x}\left( \rho U^{2}+P_{\rm c}+
P_{\rm g}\right) =0 &  & \label{mom:c}
\end{eqnarray}
 where

\begin{equation}
\label{P_c}
P_{\rm {c}}(x)=\frac{4\pi }{3}mc^{2}\int _{p_{inj}}^{\infty }\frac{p^{4}dp}{\sqrt{p^{2}+1}}f(p,x)
\end{equation}
is the pressure of the CR gas,  \( P_{\rm g} \) is the thermal
gas pressure, $\rho $ is its density. The lower boundary in the momentum space, \( p_{inj} \)
separates CRs from the thermal plasma that is not directly involved
in this formalism but rather enters the equations through the magnitude
of \( f \) at \( p=p_{inj} \) which specifies the injection rate
of thermal plasma into the acceleration process. The particle momentum
\( p \) is normalized to \( mc \).

\begin{figure}
\plotone{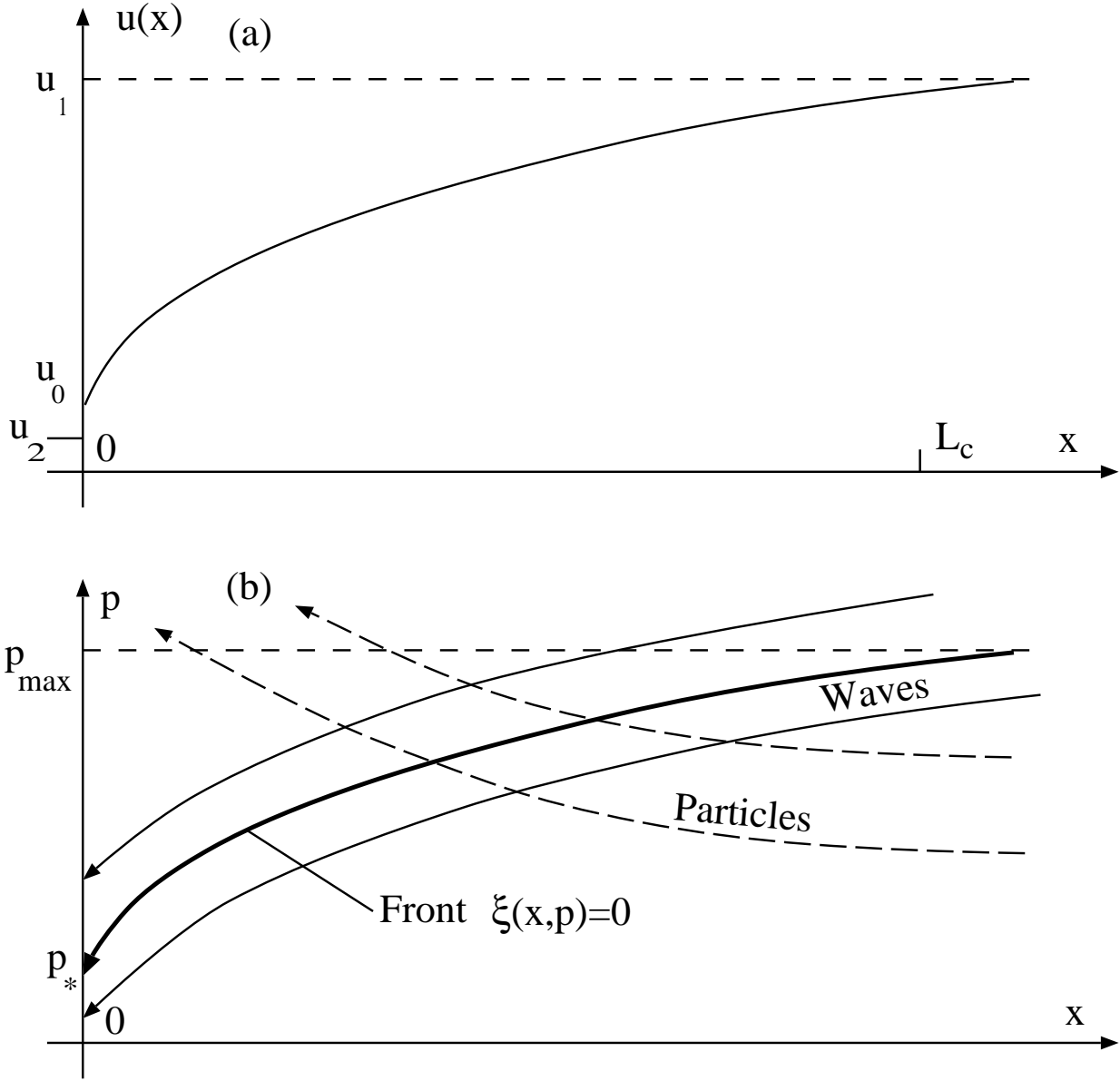}

\caption{The flow structure (a) and the phase plane of particles and resonant
waves (b).\label{fig:ph:plane}}
\end{figure}

Since we shall be primarily concerned with the wave generation and
particle confinement upstream of the discontinuity, we assume that
the upstream region is at \( x>0 \), so that the velocity profile
can be represented in the shock frame as \( U(x)=-u(x) \) where the
(positive) flow speed \( u(x) \) jumps from \( u_{2}\equiv u(0-) \)
downstream to \( u_{0}\equiv u(0+)>u_{2} \) across the sub-shock
and then gradually increases up to \( u_{1}\equiv u(+\infty )\geq u_{0} \)
(see Figure~\ref{fig:ph:plane}a below).

We may neglect the contribution of the gas pressure to eq.(\ref{mom:c})
in the upstream region (\( x>0 \), but not at \( x\le 0 \)) restricting
our consideration to the high Mach number shocks, \( M\gg 1 \). The
gas pressure is retained when treating the sub-shock (discontinuous
part of the shock structure) which, however, can be simply described
by the conventional Rankine-Hugoniot jump condition (since \( P_{c} \)
does not vary on this scale)

\begin{equation}
\label{c:r}
\frac{u_{0}}{u_{2}}=\frac{\gamma +1}{\gamma -1+2M_{0}^{-2}}
\end{equation}
 Here \( M_{0} \) is the Mach number in front of the sub-shock. When
the flow compression in the CR precursor can be considered as adiabatic,
this can be expressed through the given far upstream Mach number in
a standard way, \( M_{0}^{2}=M^{2}/R^{\gamma +1} \), where \( R\equiv u_{1}/u_{0} \)
is the flow precompression in the CR precursor. We shall also set
\( \gamma =5/3 \) in what follows.

Turning to determination of the CR diffusion coefficient \( \kappa  \),
we note that since the CR precursor scale height is \( \sim \kappa (p_{max})/u_{1}\sim (c/u_{1})r_{g}(p_{max}) \),
which is \( c/u_{1}\gg 1 \) much larger than the longest wave in
the spectrum \( \sim r_{g}(p_{max}) \) we can use a wave kinetic
equation in the eikonal approximation for describing the evolution
of Alfven waves

\begin{equation}
\label{wke}
\frac{\partial N_{k}}{\partial t}+\frac{\partial \omega }{\partial k}\frac{\partial N_{k}}{\partial x}-
\frac{\partial \omega }{\partial x}\frac{\partial N_{k}}{\partial k}=\gamma _{k}N_{k}+St\left\{ N_{k}\right\}
\end{equation}
 Here \( N_{k} \) is the number of wave quanta and \( \omega  \)
is the wave frequency \( \omega =-ku+kV_{\rm A}\simeq -ku \). The
left hand side has a usual Hamiltonian form that states the conservation
of \( N_{k} \) along the lines of constant frequency \( \omega (k,x)=\const  \)
on the \( k,x \) plane. The first term on the r.h.s. describes the
wave generation on the cyclotron instability of a slightly anisotropic
particle distribution. It can be expressed through its spatial gradient.
The resonance condition for the wave-particle interaction contains
also the particle pitch angle \( \cos ^{-1}\mu  \) by means of the
following expression \( kp\mu =eB/c \) which, generally speaking,
requires the treatment of particle distribution in two dimensional
momentum space (\( p,\mu  \)). A significant simplification can be
achieved by the so called {}``resonance sharpening{}'' procedure
(\citealt{skil75c, dru96}) whereby a certain {}``optimal{}'' value
of \( \mu  \) is ascribed to all particles and the resonance condition
puts \( k \) and \( p \) into a one-to-one relation, \ie \( kp=\const  \).
The second term on the r.h.s. stands for nonlinear wave-particle and
wave-wave interactions such as the induced scattering of waves on
thermal protons and mode coupling \citep{gs69}. We will suggest
a simple model for this nonlinear term in section \ref{sec:modif}. 

To conclude this subsection we emphasize that while eqs.(\ref{dc1}-\ref{mom:c})
already treat the acceleration process and flow structure on equal
footing, the fluctuation part given by eq.(\ref{wke}) must be included
in this treatment and, as we shall see in the sequel, it by no means
plays a subdominant role in this triad.

\subsection{The significance of acceleration nonlinearities\label{sec:signif}}

There are two aspects of the acceleration where nonlinearity is crucial
for its outcome. The first aspect is the excitation of scattering
waves by accelerated particles and the second one is the backreaction
of these particles on the shock structure. The latter is critical
both for the particle injection and wave excitation that is, to particle
confinement. 

Indeed, the system \ref{dc1}-\ref{P_c} self-consistently describes
particle acceleration and the shock structure (nonlinearly modified
by the particle pressure) only if the particle scattering law is known
(which is contained in the diffusion coefficient \( \kappa  \)) and
the injection rate from the thermal plasma is also known (the normalization
of the particle distribution \( f(p) \) in eq.(\ref{dc1})). Physically,
the scattering rate determines the particle maximum momentum \( p_{max} \),
as eq.(\ref{t:acc}) indicates. The difficulty, however, is that both
the cut-off momentum \( p_{max} \) and the wavenumber cutoff of the
scattering turbulence change in time \emph{simultaneously} (one controlling
the other) due to the cyclotron resonance condition. However the speed
at which they change has not been calculated self-consistently. The
\emph{linear} solution given by eqs.(\ref{p:max}) and (\ref{t:acc})
is essentially based on the assumption that \( p_{max} \) is growing
due to \emph{already existing stationary turbulence}. In reality,
the particle energy cut-off and the corresponding cut-off on the wave
spectrum, as we mentioned, both advance together and, since waves
need to grow from a very small background amplitude at each current
cut-off position, an additional slow down must be introduced in the
entire process. A good analogy here is the problem of beam relaxation
in plasmas \citep{ivanov} (where a front on the particle velocity
distribution also propagates on self-generated rather than on pre-existing
resonant waves). This suggests that the speed of the front in momentum
space, as given by eqs.(\ref{p:max}) and (\ref{t:acc}) should be
reduced by a factor \( \sim \ln (W/W_{ISM}) \), where \( W_{ISM} \)
is the background turbulence amplitude and \( W \) is the saturated
wave amplitude generated by accelerated particles. As we mentioned,
the latter may be associated with \( W_{B}=B^{2}_{0}/8\pi  \) so
that the acceleration time given by eq.(\ref{t:acc}) may increase
by a factor \( \sim 10 \) (\eg \citealt{abr94} estimate \( W_{ISM}/W_{B}\sim 10^{-5} \)).
Evidently, the additional \( ln \)-factor takes care of the time
needed for waves to grow before they start to scatter particles with
current momentum \( p \) at the Bohm rate.

The above consideration also shows that particles with \( p<p_{max} \)
are confined to the shock through fast pitch angle scattering while
particles with \( p>p_{max} \) are scattered only very slowly due
to the absence of self-generated waves and leave the accelerator.
Mathematically, this means \( f(p>p_{max})\equiv 0 \) or \( \kappa (p>p_{max})\equiv \infty  \).
Note that the propagating front solution must produce different (sharper)
cut-off shape at \( p=p_{max}(t) \) than approaches based on the
pre-existing turbulence, \ie on a prescribed (for all \( p) \) \( \kappa (p) \),
(\eg \citealt{byk96}). Even if the speed and the form of the front at \( p_{max}(t) \)
are unknown, the above ansatz allows analytic solution of the system
(\ref{dc1}-\ref{P_c}) \citep{m97a} for \( p<p_{max} \) in the limit
of strong shocks \( M\gg 1 \), high maximum momentum \( p_{max} \)
(that may slowly advance in time) and for essentially arbitrary, in
particular, Bohm \( \kappa (p) \) dependence for \( p<p_{max} \)
(as we mentioned, often assumed in numerical studies, however, for
all \( p \), \eg \citealt{DuffyB}). The analytic solutions are tabulated
\eg by \cite{mdru01} and extensively used below.

Since waves are generated by accelerated particles upstream in the
precursor, the main nonlinear impact on the wave dynamics and thus
on \( p_{max} \) must be from the flow pre-compression. The latter
can be characterized by the parameter \( R=u_{1}/u_{0} \) which is
shown in Figure~\ref{fig:bif} as a function of injection parameter \( \nu  \)
for different maximum momenta \( p_{max} \). The injection parameter
\( \nu  \) is related to the normalization of particle distribution
function \( f \) in eq.(\ref{dc1}) as 

\begin{equation}
\label{inj:def}
\nu =\frac{4\pi }{3}\frac{mc^{2}}{\rho _{1}u^{2}_{1}}p_{inj}^{4}f_{0}(p_{inj})
\end{equation}
where \( f_{0}(p) \) is the downstream value of \( f \). In this
form the injection rate \( \nu  \) naturally appears as a coefficient
in front of the CR pressure in the momentum flux conservation eq.(\ref{mom:c})
when the CR pressure is normalized to the ram pressure \( \rho _{1}u_{1}^{2} \)
(see eq.{[}\ref{Pc:norm}{]} in the next section).

One aspect of the solution shown in Figure~\ref{fig:bif} which is important
here is that for any given injection rate \( \nu  \), the growing
maximum momentum \( p_{max}(t) \) will ultimately exceed a critical
value, beyond which the test particle regime fails to exist. (It is
natural to assume that the acceleration starts at this regime \ie where
\( R\approx 1 \), \eg point A on Figure~\ref{fig:bif}). Formally, the
system must then transit to a much higher \( R \) that will be still
very sensitive to the current values of \( \nu (\approx 0.01)  \) and \( p_{max} (=10^6) \)
as may be seen from Figure~\ref{fig:bif} (point B). Obviously, the further
development of the acceleration process will depend on how the parameters
\( \nu  \) and \( p_{max} \) react to this strong increase of \( R \).
One possibility is to assume that simply a constant fraction of the
sub-shock plasma is injected so that the injection rate substantially
increases because the plasma density at the sub-shock grows linearly
with \( R \). Then, the system must leave the critical region where
the \( R(\nu ) \) dependence is very sharp or even nonunique and
proceed to a highly supercritical regime characterized by higher \( \nu  \)
(point C). The curve \( R(\nu ) \) saturates there at the level \( \propto M^{3/4} \)
which in the most straightforward way may be deduced from the condition
of the sub-shock preservation, \( M_{0}\ga 1 \), (see eq.[\ref{c:r}]).
A general formula for \( R(\nu ,M) \) with the \( M^{3/4} \) scaling
as a limiting case may be found in \citep{m97a}. This scenario was
realized in many numerical models (\eg \citealt{el:eich85,kazel86,byk96}),
since they normalized the injection parameter to the
plasma density at the sub-shock \( \rho _{0}=\rho _{1}R \), which
should clearly lead to the \( R\propto M^{3/4} \) scaling.  Obviously,
the pre-compression \( R \) and thus the acceleration efficiency
will then be insensitive to \( \nu  \) (in deep contrast to the
case $\nu\approx=const$, point B) since the point \( C \) is
already on the saturated part of the \( R(\nu ) \) curve. Often,
this insensitivity is observed in numerical studies with the parameterized
injection rate (\eg \citealt{byk96}), so it is tempted to conclude that
we do not need to know the injection rate very accurately, as soon
as it exceeds the critical rate.

However, the injection rate is known to be suppressed by a number
of self-regulating mechanisms such as trapping of thermal particles
downstream by the injection driven turbulence  \citep{m98} and cooling
of the downstream plasma in strongly modified shocks. These effects
are believed to more than compensate the compressive growth of plasma
density. Recently, these effects have been systematically included
in numerical studies by \cite{gjk00, kjg01}. They did not confirm
the simple \( \nu \propto R \) rule. Instead they indicate that in
course of nonlinear shock modification accompanied by growing \( R \),
the injection rate \( \nu  \) remains remarkably constant \citep{gjk00}.
Moreover, the preliminary results of a new adaptive mesh refinement
(AMR) modification of these scheme allowing higher \( p_{max} \),
indicate that the injection efficiency may even begin to decrease
with growing \( p_{max} \) (\citealt{kjg01, kjg02}). These self-regulation
mechanisms are applicable to both strictly parallel and oblique shocks
of which the former ones is clearly an exceptional case. Even slightly
oblique shocks have an additional self-regulation of injection via
a nonlinearly increasing obliquity. Indeed, since the tangential magnetic
field component \( B_{t} \) is amplified at the sub-shock by the
factor of \( R \), the sub-shock may be strongly oblique even if
the shock itself is not. This leads to exponentially strong suppression
of leakage of downstream thermal particles upstream (for a Maxwellian
downstream distribution) since the intersection point of a field line
(the particles sit on) with the shock front rapidly escapes from these
particles. On the other hand, enhanced particle reflection off
the oblique subshock should increase injection.

The inspection of Figure~\ref{fig:bif} shows that if we (conservatively)
assume \( \nu (R)=const \) (AB) rather than \( \nu \propto R \)
(AC) the results will differ dramatically particularly in terms of
the injection rate. Note that particle spectra that correspond to
the points \( A \) and \( C \) also differs very strongly (see \citealt{mdru01}
for graphical examples). What is important for the subject of the
present paper is that in both these cases as well as for any other
point on the part \( BC \) of the bifurcation curve, the compression
\( R \) is very high. It has been pointed out by \cite{mdv00}, that
this must have strong impact not only on the injection rate as discussed
above, but also on the wave propagation and thus on the particle confinement.
This in turn should lead to significant reduction of the maximum momentum
achievable by this acceleration mechanism. We will quantify these
effects in the next section.

\section{Analysis}

Returning to eqs.(\ref{dc1}) and (\ref{wke}), it is convenient to
use the wave energy density normalized to \( d\ln k \) and to the
energy density of the background magnetic field \( B^{2}_{0}/8\pi  \)
instead of \( N_{k} \) 

\begin{equation}
\label{en:dens}
I_{k}=\frac{k^{2}V_{\rm A}}{B^{2}_{0}/8\pi }N_{k}
\end{equation}
 along with the partial pressure of CRs normalized to \( d\ln p \)
and to the shock ram pressure \( \rho _{1}u^{2}_{1} \)

\begin{equation}
\label{Pc:norm}
P=\frac{4\pi }{3}\frac{mc^{2}}{\rho _{1}u^{2}_{1}}\frac{p^{5}}{\sqrt{p^{2}+1}}f(p,x)
\end{equation}
 Using these variables, denoting \( g=P/p \), assuming a steady state
and \( p\gg 1 \), eqs.(\ref{dc1},\ref{wke}) can be rewritten as
(\citealt{bell78a, dru96})

\begin{eqnarray}
\frac{\partial }{\partial x}\left( ug+\kappa \frac{\partial g}{\partial x}\right) =
\frac{1}{3}u_{x}p\frac{\partial g}{\partial p} &  & \label{dc2} \\
u\frac{\partial I}{\partial x}+u_{x}p^{3}\frac{\partial }{\partial p}\frac{I}{p^{2}}=
\frac{2u_{1}^{2}}{V_{\rm A}}\frac{\partial }{\partial x}P-St\left\{ I\right\}  &  & \label{wke2}
\end{eqnarray}
Here \( u_{x}\equiv \partial u/\partial x \) and the wave intensity
\( I\equiv I(p)=I_{k} \) is now treated as a funcion of \( p \)
rather than \( k \) according to the resonance relation \( kp=const \).
The CR diffusion coefficient \( \kappa  \) can be expressed through
the wave intensity by

\begin{equation}
\label{kappa}
\kappa =\frac{\kappa _{\rm B}}{I}
\end{equation}
 where \( \kappa _{\rm B}(p) \) is the Bohm diffusion coefficient.
The difference between these equations and those used by, \eg \cite{bell78a, dru96}
is due to the terms with \( u_{x}\neq 0 \) and the \( St \)-term
on the r.h.s. of eq.(\ref{wke2}). Far away from the sub-shock where
\( u_{x}\to 0 \), and where the wave collision term is also small
due to the low particle pressure \( P \), one simply obtains

\begin{equation}
\label{I:lin}
I=\frac{2u_{1}}{V_{\rm A}}P
\end{equation}
 Note, that this shows the limitation of the linear approach in the
case of strong shocks \( M_{\rm A}\equiv u_{1}/V_{\rm A}\gg 1 \).
The most important change to the acceleration process comes from the
terms with \( u_{x}\neq 0 \). Indeed, let us recall first how the
equation (\ref{dc2}) may be treated in the linear case \( u_{x}\equiv 0 \)
for \( x>0 \). We integrate both sides between some \( x>0 \) and
\( +\infty  \), which yields

\begin{equation}
\label{cd:lin}
u_{1}g+\frac{\kappa _{0}V_{\rm A}}{2u_{1}}\frac{1}{g}\frac{\partial g}{\partial x}=0
\end{equation}
 where we denoted \( \kappa _{0}\equiv \kappa _{\rm B}/p\simeq const \)
for \( p\gg 1 \). Although this equation has a formal spatial scale
\( l\sim \kappa _{0}/u_{1}M_{\rm A}g \), its only solution is a power
law 

\begin{equation}
\label{g:bell}
g\propto 1/\left( x+x_{0}\right) 
\end{equation}
 and thus has no scale (\( x_{0}=x_{0}(p) \) is an integration constant).
It simply states the balance between the diffusive flux of particles
upstream (second term in eq.{[}\ref{cd:lin}{]}) and their
advection with thermal plasma downstream (the first term). As we shall
see, this balance is possible not everywhere upstream and the physical
reason why it appears to be so robust in the case \( u_{x}=0 \) is
that flows of particles and waves on the \( x,p \)-plane (including
the diffusive particle transport) are both directed along the \( x \)-axis.
If, however, the flow modification upstream is significant (\( u_{x}>0 \),
\( x>0 \)), the situation changes fundamentally. Figure~\ref{fig:ph:plane}
explains how the flows of particles and waves on the \( x,p \)-plane
become misaligned even though they are both advected with the thermal
plasma. In fact, the flows separate from each other and, since neither
of them can exist without the other (waves are generated by particles
that, in turn, are trapped in the shock precursor by the waves) they
both disappear in some part of the phase space. To understand how
this happens we rewrite eqs.(\ref{dc2}-\ref{wke2}) in the following
characteristic form (we return to the particle number density \( f \))

\begin{eqnarray}
\left( u\frac{\partial }{\partial x}-\frac{1}{3}u_{x}p\frac{\partial }{\partial p}\right) f=
-\frac{\partial }{\partial x}\kappa \frac{\partial f}{\partial x} &  & \label{dc3} \\
\left( u\frac{\partial }{\partial x}+u_{x}p\frac{\partial }{\partial p}\right) \frac{I}{p^{2}}=
\frac{2u_{1}^{2}}{V_{\rm A}p^{2}}\frac{\partial }{\partial x}P-\frac{1}{p^{2}}St\left\{ I\right\}  &  & \label{wke3}
\end{eqnarray}
 One sees from the l.h.s.'s of these equations that particles are
transported towards the sub-shock in \( x \) and upwards in \( p \)
along the family of characteristics \( up^{3}=const \), whereas waves
move also towards the sub-shock but downwards in \( p \) along the
characteristics \( u/p=const \). As long as \( u(x) \) does not
significantly change the waves and particles propagate together (along
\( x \)-axis) as \eg in the case of unmodified shock or far away
from the sub-shock where \( u_{x}\rightarrow 0 \). When the flow
compression becomes important (\( u_{x}\neq 0 \)) their separation
leads to decrease of both the particle and wave energy densities towards
the sub-shock. To describe this mathematically, let us assume first
that the relation (\ref{I:lin}) between \( P \) and \( I \) is
still a reasonable approximation even if \( u_{x} \) is nonzero but
small. Then, integrating eq.(\ref{dc2}) again between some \( x>0 \)
and \( x=\infty  \), instead of (\ref{cd:lin}) we obtain

\begin{equation}
\label{cd:int}
ug+u_{1}\frac{L}{g}\frac{\partial g}{\partial x}=-\frac{1}{3}\int _{x}^{\infty }u_{x}p\frac{\partial g}{\partial p}dx
\end{equation}
 In contrast to the solution of eq.(\ref{cd:lin}) the length scale
\( L\equiv \kappa _{0}/2u_{1}M_{\rm A} \) enters the solution of
this equation. This is because it has a nonzero r.h.s. In the next
subsection we obtain the solution of this equation that rapidly changes
on a scale \( \sim L \).

\subsection{Internal asymptotic solution for \protect\( g\protect \) \label{sec: int} }

First we note that \( L\ll L_{\rm c} \) where \( L_{\rm c}=\kappa (p_{\rm max})/u_{1} \)
is the total scale height of the CR precursor on which \( u(x) \)
changes. Next, in addition to \( x \) and \( p \), we introduce
a fast (internal) variable \( \xi (x,p) \) as follows

\begin{equation}
\label{ksi}
\xi =\frac{x-x_{\rm f}(p)}{L}
\end{equation}
 where \( x=x_{\rm f}(p) \) is some special curve on the \( x,p \)
plane which bounds the solution and will be specified later. We rewrite
eq.(\ref{cd:int}) for

\begin{equation}
\label{order}
\xi \; -{\rm fixed},\; L\to 0
\end{equation}
Separating fast variable terms on the r.h.s. by replacing 

\[
\partial g/\partial p\to \partial g/\partial p-L^{-1}\left( \partial x_{\rm f}/\partial p\right) \left( \partial g/\partial \xi \right) ,\]
 to the leading order in \( L/L_{\rm c}\to 0 \), we obtain

\begin{equation}
\label{dc:int2}
u_{\rm f}g+\frac{u_{1}}{g}\frac{\partial g}{\partial \xi }=\frac{1}{3}p\frac{du_{\rm f}}{dp}\left( G-g\right) -
\frac{1}{3}\int _{x}^{\infty }u_{x}p\frac{\partial G}{\partial p}dx
\end{equation}
Here we denoted \( u_{\rm f}(p)\equiv u\left[ x_{\rm f}(p)\right]  \)
and 

\begin{equation}
\label{lim:g}
G(x,p)=\lim _{\xi \to \infty }g(\xi ,x,p)
\end{equation}
The existence of this limit will be confirmed upon obtaining the solution
of eq.(\ref{dc:int2}) below. First, we introduce the following notations

\begin{eqnarray*}
w(p)=u_{\rm f}+\frac{1}{3}p\frac{du_{\rm f}}{dp} &  & \\
S(x,p)=\frac{1}{3}p\frac{du_{\rm f}}{dp}G-\frac{1}{3}\int _{x}^{\infty }u_{x}p\frac{\partial G}{\partial p}dx &  &
\end{eqnarray*}
 Eq.(\ref{dc:int2}) then can be rewritten as: 

\begin{equation}
\label{dc:int3}
wg+\frac{u_{1}}{g}\frac{\partial g}{\partial \xi }=S
\end{equation}
 and its solution can be thus written as 

\begin{equation}
\label{g:intern}
g(\xi ,x,p)=\frac{S(x,p)}{w(p)+e^{-S\xi /u_{1}}}
\end{equation}
One sees that the limit in eq.(\ref{lim:g}) indeed exists and is
equal to \( G=S/w \). Furthermore, eq.(\ref{g:intern}) describes
a transition front on the particle distribution between its asymptotic
value \( g=G \) at \( \xi \to \infty  \) and \( g=0 \) at \( \xi \to -\infty  \).
This front solution establishes as a result of particle losses caused
by the lack of resonant waves towards the sub-shock as we argued discussing
eqs.(\ref{dc3},\ref{wke3}). Note that according to the ordering
in eq.(\ref{order}) we should set \( x=x_{\rm f}(p) \) in \( S(x,p) \)
when solving (\ref{dc:int3}) for \( g(\xi ) \) and we indeed must
do it for \( \xi \sim 1 \) as well as for all negative \( \xi <0 \).
In the limit \( \xi \to \infty  \), however, we may use the result
(\ref{g:intern}) for arbitrary \( x>x_{\rm f}(p) \) since it remains
valid in this case, however, it merely states that in this region
the complete solution is represented by its {}``external{}'' part
\( G(x,p) \) (eq.{[}\ref{lim:g}{]}). This, in turn, is yet to be
determined. Before we do this in section~\ref{sec: ext} we should verify
the validity of the internal solution and calculate its unknown function
\( x_{\rm f}(p) \).

\subsection{Nonlinear modification of the internal solution. Determination of
\protect\( x_{\rm f}(p)\protect \)\label{sec:modif}}

The way we resolved eq.(\ref{wke2}) for \( I \) (see eq.(\ref{I:lin}))
may become inadequate for two reasons. First, the second term on the
l.h.s. of eq.(\ref{wke2}) may become comparable with the first one.
This problem could be resolved, however, by integrating this equation
along the characteristic \( u/p=const \) instead of the \( x \)-axis
as we did to obtain eq.(\ref{I:lin}). The matter of bigger concern
is that the increase of \( u_{x} \) is obviously has to do with strong
shock modification, so that \( P\sim 1 \). Clearly, under these circumstances
the balance between the l.h.s. of eq.(\ref{wke2}) and the pressure
term on the r.h.s. leads to impossibly large \( I \). Evidently,
the second term on the r.h.s. must come into play before this has
happened so that the steady state will be maintained by the balance
between this term and the pressure term while the l.h.s. will become
sub-dominant. Thus, for \( I \) we have the following equation

\begin{equation}
\label{bal:nl}
\frac{2u_{1}^{2}}{V_{\rm A}}\frac{\partial }{\partial x}P-St\left\{ I\right\} =0
\end{equation}

As it is often the case we may assume that the wave collision term
\( St\{I\}\propto I^{2} \) and in the long wave limit \( k\to 0 \)
it is also proportional to \( k^{2} \) (that means to \( p^{-2} \)).
The pressure gradient may be estimated as \( P/L_{p} \) where \( L_{p} \)
is the scale height of particles of momentum \( p \) which we assume
for simplicity to be proportional to \( p \) as in the standard Bohm
case. Thus, for \( I \) we have

\begin{equation}
\label{I:nlin}
I^{2}\simeq \frac{u_{1}}{\alpha V_{\rm A}}pP
\end{equation}
where \( \alpha  \) characterizes the strength of nonlinear wave
interaction. Using the last estimate, instead of eq.(\ref{cd:int})
we have

\begin{equation}
\label{cd:int2}
ug+u_{1}\frac{L_{\rm nl}}{\sqrt{g}}\frac{\partial g}{\partial x}=
-\frac{1}{3}\int _{x}^{\infty }u_{x}p\frac{\partial g}{\partial p}dx
\end{equation}
 where \( L_{\rm nl}=\kappa _{0}/u_{1}\sqrt{\alpha M_{_{\rm A}}} \).
Introducing the fast variable \( \xi  \) (\ref{ksi}) with \( L_{\rm nl} \)
instead of \( L \), and repeating the derivation in section~\ref{sec: int},
for \( g \) we obtain the following equation 

\[
wg+\frac{u_{1}}{\sqrt{g}}\frac{\partial g}{\partial \xi }=S\]
with the obvious solution 

\begin{equation}
\label{g:nlin}
g=\frac{S}{w}\tanh ^{2}\frac{\sqrt{wS}}{2u_{1}}\xi 
\end{equation}
This solution is valid for \( \xi \ga \xi _{0}>0 \) (\( \xi _{0}\sim G^{-1} \))
whereas at \( -\infty <\xi \la \xi _{0} \) one should use the linear
formula (\ref{I:lin}) for the wave spectral density and thus the
solution (\ref{g:intern}) instead of eq.(\ref{g:nlin}). The uniformly
valid solution may be also obtained by using an interpolation between
eq.(\ref{I:lin}) and eq.(\ref{I:nlin}) for \( I \). We will not
need, however, the explicit form of the front transition in the particle
distribution in the region \( \xi \sim 1 \) which means \( x\approx x_{\rm f}(p) \).
We will merely exploit the fact that this transition is much narrower
(its width is \( \Delta x\sim L_{\rm nl}/\sqrt{G(x_{\rm f},p)} \))
than that of the main part \( G(x) \) in the interval \( x_{\rm f}<x<\infty  \).
The spatial scale of the latter is at least \( \sim \kappa _{\rm B}/u_{1} \)
or even broader if the linear approximation (\ref{g:bell}) can be
used, in which case the length scale is determined by the linear damping
of Alfven waves \citep{dru96}.

The only characteristic of the above internal solution that is needed
to calculate the external solution \( G(x,p) \) is the position of
the front transition in \( g \) on the \( x,p \)-plane, \ie we need
to calculate the function \( x=x_{\rm f}(p) \). To do this we return
to eq.(\ref{wke3}). We solved it by neglecting its l.h.s. and finally
arrived at the result for \( g \) and thus for \( I \) in eq.(\ref{wke3})
that contains the fast variable \( \xi  \). Generally, this produces
large terms in the next order of approximation coming from the l.h.s.
To avoid that we must choose the position of the transition front
(\( \xi (x,p)=0 \)) in such a way that it coincides with one of the
characteristics of the operator on the l.h.s. of eq.(\ref{wke3}),
\ie

\[
\left( u\frac{\partial }{\partial x}+u_{x}p\frac{\partial }{\partial p}\right) \xi (x,p)=0\]
or 

\[
u_{\rm f}(p)-p\frac{du_{\rm f}}{dp}=0\]
The choice of the concrete characteristic is based on the existence
of the absolute maximum momentum \( p_{\rm max} \) beyond which there
are neither particles nor waves. That means

\[
u_{\rm f}(p)\equiv u\left[ x_{\rm f}(p)\right] =u_{1}\frac{p}{p_{\rm max}}\]
Likewise, the function \( x=x_{\rm f}(p) \) is defined as

\[
x_{\rm f}(p)=u^{-1}\left( u_{1}\frac{p}{p_{\rm max}}\right) \]

\subsection{External solution\label{sec: ext}}

While having obtained the form and the position \( x=x_{\rm f}(p) \)
of the narrow front in the particle distribution \( g(x,p) \) we
still need to calculate \( g \) to the right from the front where
it decays with \( x \). This would be the external solution \( G(x,p) \)
introduced in the previous subsections. It is clear that 

\[
\max _{x}g(x,p)\approx G(x_{\rm f},p)\equiv G_{0}(p)\]
so that from eq.(\ref{cd:int}) we have the following equation 

\begin{equation}
\label{cd:ext}
u_{\rm f}(p)G_{0}(p)=-\frac{1}{3}\int _{x_{\rm f}(p)}^{\infty }u_{x}p\frac{\partial G}{\partial p}dx
\end{equation}
The most important information about \( G(x,p) \) is contained in
\( G_{0}(p) \) for which from the last equation we obtain

\begin{equation}
\label{G0:eq}
\frac{\partial }{\partial p}v(p)G_{0}(p)+4\frac{u_{1}}{p_{1}}G_{0}(p)=0
\end{equation}
where we have introduced \( v(p) \) by

\begin{equation}
\label{v:def}
v(p)=\frac{1}{G_{0}(p)}\int ^{u_{1}}_{u_{\rm f}(p)}G(x,p)du(x)
\end{equation}
 Eq.(\ref{G0:eq}) can be easily solved for \( G_{0} \) 

\begin{equation}
\label{G0:sol}
G_{0}(p)=\frac{C}{v(p)}\exp \left( -4\frac{u_{1}}{p_{\max }}\int \frac{dp}{v(p)}\right) 
\end{equation}
 where \( C \) is a normalization constant that should be determined
from matching this solution with that in the region \( p<p_{*} \).
However, the function \( v \) depends on the solution itself. Fortunately,
this quantity can be calculated prior to determining \( G_{0} \)
and therefore, this solution may be written in a closed form. To illustrate
this, let us consider a particularly simple case of \( p\simeq p_{\max } \),
and we will turn to the general case afterwards. Clearly, \( p\simeq p_{max} \)
means \( u_{\rm f}(p)\simeq u_{1} \). Evidently, we may replace \( G \)
in eq.(\ref{v:def}) by \( G_{0} \) so that for \( \nu (p) \) we
have \( v(p)\simeq u_{1}-u_{\rm f}(p)=u_{1}\left( 1-p/p_{\max }\right)  \).
Thus, from eq.(\ref{G0:sol}) we obtain the following shape of the
cut-off near \( p_{max} \)

\begin{equation}
\label{G0:pmax}
G_{0}\simeq C\left( p_{\max }-p\right) ^{3}
\end{equation}
In the rest of the \( x,p \)-domain where \( x>x_{\rm f}(p) \) and
\( p \) is not close to \( p_{\max } \), we may assume that the
CR diffusion coefficient is close to its Bohm value.  Indeed, in contrast
to the phase space region \( x\approx x_{\rm f}(p) \) at each given
\( x,p \) there are waves generated along the entire characteristic
of eq.(\ref{wke2}) passing through this point of the phase space
and occupying an extended region of the CR precursor, Figure~\ref{fig:ph:plane}.
We may use then the asymptotic high Mach number solution found in
\citep{m97a}

\begin{equation}
\label{g:xp}
g(x,p)=g_{0}(p)\exp \left( -\frac{1+\beta }{\kappa (p)}\int ^{x}_{0}udx\right) 
\end{equation}
Here \( \beta  \) is numerically small (typically \( \simeq 1/6 \))
and this solution without \( \beta  \)-term manifests the balance
between the diffusion and convection terms on the l.h.s. of eq.(\ref{dc2})
which is more accurate approximation far upstream where the flow modification
(r.h.s.) is weak. The flow profile depends on the form of \( \kappa (p) \)
and for \( \kappa =\kappa _{\rm B}\propto p \) in the internal part
of the shock transition \( u(x) \) behaves linearly with \( x \).
Adopting this solution to the region \( x>x_{\rm f} \), we may write

\[
G(x,p)=G_{0}(p)\exp \left( -\frac{1+\beta }{\kappa _{\rm B}}\int ^{x}_{x_{\rm f}}udx\right) \]
so that for \( v \) we have

\[
v(p)=\int ^{u_{1}}_{u_{\rm f}(p)}du\exp \left( -\frac{1+\beta }
{\kappa _{\rm B}}\int ^{u}_{u_{\rm f}}\frac{u'du'}{u_{x}(u')}\right) \]
In the Bohm case we can use the simplified linear approximation for
\( u(x) \) \cite{m97a}, \( u=u_{0}+u_{1}x/L_{\rm c} \) where \( L_{\rm c}=
\pi \kappa _{\rm B}(\hat{p})/2\theta u_{1} \),
\( \theta \approx 1.09 \) and it is implied that the maximum contribution
to the particle pressure comes from the momentum \( p=\hat{p} \)
(specified later). Now we can express \( v \) in the form of an error
integral

\begin{equation}
\label{nu:gen}
v(p)\simeq \int ^{u_{1}}_{u_{\rm f}}du\exp \left[ -\frac{\left( 1+\beta \right) L_{\rm c}}{2u_{1}
\kappa _{\rm B}(p)}\left( u^{2}-u^{2}_{\rm f}\right) \right]
\end{equation}
The algebra further simplifies in two limiting cases (the second of
which has been already mentioned)

\[
v(p)=u_{1}\left\{ \begin{array}{cc}
\sqrt{\pi u_{1}\kappa _{\rm B}/2\left( 1+\beta \right) L_{\rm c}}, & p\ll p_{\max }\\
\left( 1-p/p_{\max }\right) , & p\simeq p_{\max }
\end{array}\right. \]
This yields the following asymptotic behaviour of \( G_{0}(p) \)

\begin{equation}
\label{G0:sol1}
G_{0}(p)=C\left\{ \begin{array}{cc}
\sqrt{\hat{p}\left( 1+\beta \right) /\theta p}\exp \left( -\frac{8}{p_{max}}\sqrt{\left( 1+\beta \right) p\hat{p}/\theta }\right) , & p\ll p_{\max }\\
\left( p_{\max }-p\right) ^{3} & p\simeq p_{\max }
\end{array}\right. 
\end{equation}
This result was obtained for particles with momenta \( p\ge p_{*}\equiv p_{\max }/R=p_{\max }u_{0}/u_{1} \),
whereas for \( p<p_{*} \) we may use the spectra tabulated in \citep{mdru01}
for different \( p_{max} \), that should be associated now with \( p=\hat{p} \)
. The matching of these two spectra should give the normalization
constant \( C \) in the solution (\ref{G0:sol1}). This will be the
subject of the next section.

\subsection{Connection with the main part of the spectrum\label{sec:connect}}

A typical solution of the nonlinear acceleration problem with a prescribed
maximum momentum \( p_{max} \) calculated using the method of integral
equations developed in  \citep{m97a, mdru01} is shown in Figure~\ref{fig:match}
with the dash-dotted line. Since the influence of shock modification
on the injection rate is not known for high \( p_{max} \) (see, however,
\citealt{kjg01}, where the values of \( p_{max}\sim 10 \) have been
reached) we have taken the injection rate \( \nu \approx 0.1 \),
\ie well inside the interval between the points A and C on Figure~\ref{fig:bif}
(see section~\ref{sec:signif}).

To calculate an integral spectrum containing both the part modified
by the wave compression in the shock precursor as well as its lower
energy downstream part (\( p\la p_{*} \)) we proceed as follows (see
eq.(\ref{g:xp}) for the spatial structure of the spectrum). In the
momentum range \( p<p_{*}=p_{max}/R \) we can obviously use the same
method of integral equations. However, the role of maximum momentum
is now played not by \( p_{max} \) but by \( \hat{p} \) (\ie a dynamical
cut-off where the maximum contribution to the CR pressure is coming
from). Furthermore, given \( \nu  \) and \( \hat{p} \), we calculate
the self-consistent flow structure with the precompression \( R \)
shown in Figure~\ref{fig:bif} as well as the particle spectrum. The
latter is shown in Figure~\ref{fig:match} with the dashed line. Note,
that the spectrum matching point \( p_{*} \) with the cut-off area
\( p_{*}<p<p_{max} \) is now determined and we are ready to obtain
the final spectrum by calculating its cut-off part from eqs.(\ref{G0:sol},\ref{G0:sol1},\ref{nu:gen}).
The spectrum is drawn with the full line. The momentum \( \hat{p} \)
may be obtained now as a point of maximum of the function \( pG_{0} \)
(particle partial pressure per logarithmic interval) from eq.(\ref{G0:sol1})
(upper line) which yields 

\[
\hat{p}\approx \frac{p_{max}}{8}\sqrt{\frac{\theta }{1+\beta }}\approx 0.1p_{max}\]
It should be clear from our treatment that this formula is valid only
when the shock is strongly modified, namely when \( p_{*}\equiv p_{max}/R<0.1p_{max} \)
(the case we are interested in). Since \( R \) cannot exceed \( M^{3/4} \),
this means that the shock must be sufficiently strong, \( M^{3/4}>10 \).

\section{Discussion}

\begin{figure}
\plotone{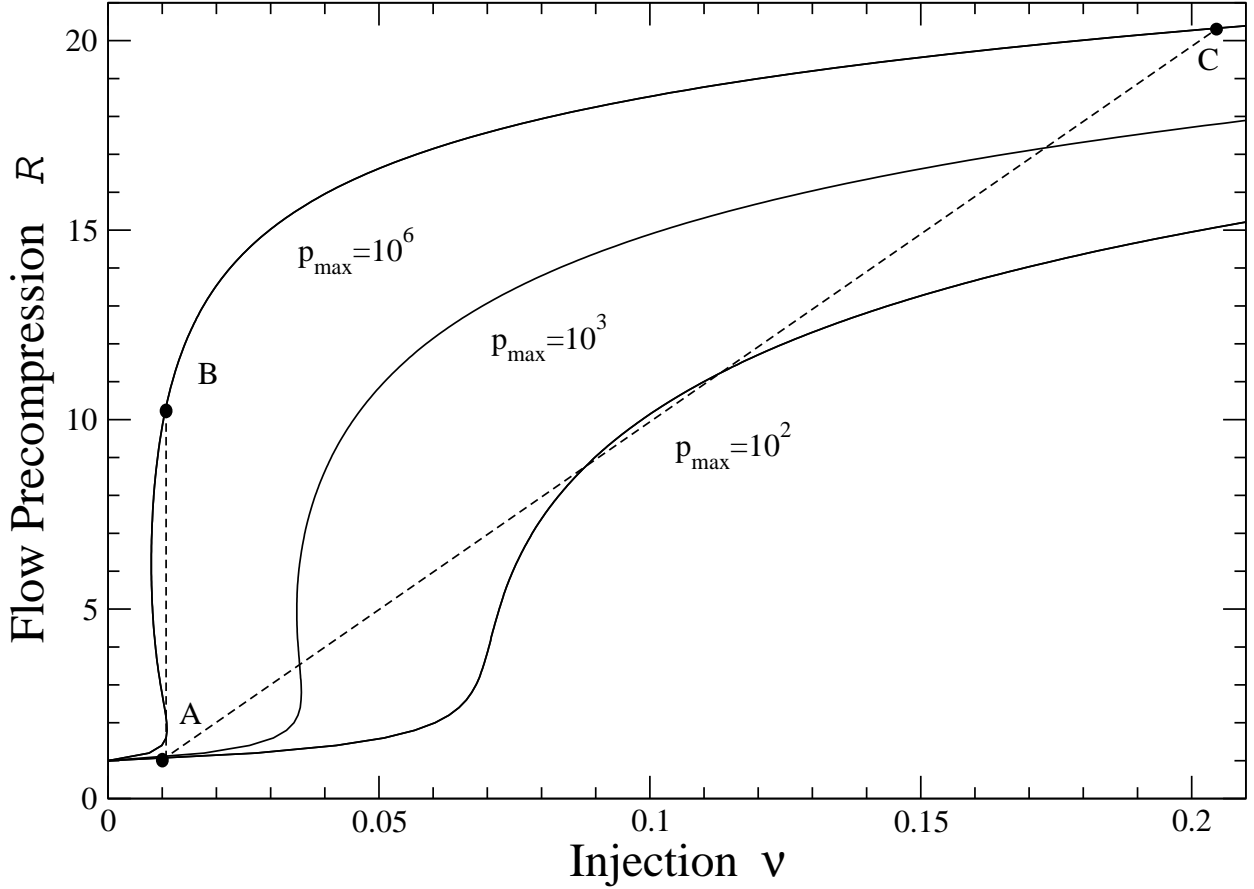}

\caption{Response of the shock structure (bifurcation diagram) to the injection
of thermal particles at the rate \protect\( \nu \protect \). The
strength of the response is characterized by the precompression of
the flow in the CR shock precursor \protect\( R=u_{1}/u_{0}\protect \).
The flow Mach number \protect\( M=150\protect \). Different curves
correspond to different values of maximum momentum normalized to \protect\( mc\protect \).
For each given \protect\( \nu \protect \) and \protect\( p_{max}\protect \),
one (for \protect\( p_{max}<p_{cr}\simeq 500)\protect \) or three
(for \protect\( p_{max}\ge p_{cr}\protect \)) solutions exist. Note
that solution multiplicity does not exist for shocks with \protect\( M\le M_{cr}\simeq 70\protect \)
\citep{mdv00, mdru01}. Given an initial injection \protect\( \nu \protect \)
and compression \protect\( R\protect \) at point \emph{A} (with \protect\( R(A)\approx 1\protect \)),
the injection and \protect\( R\protect \) at point \emph{C} is calculated
as \protect\( \nu (C)=R(C)\nu (A)\protect \) (see text for further
explanations).\label{fig:bif} }
\end{figure}

\begin{figure}
\plotone{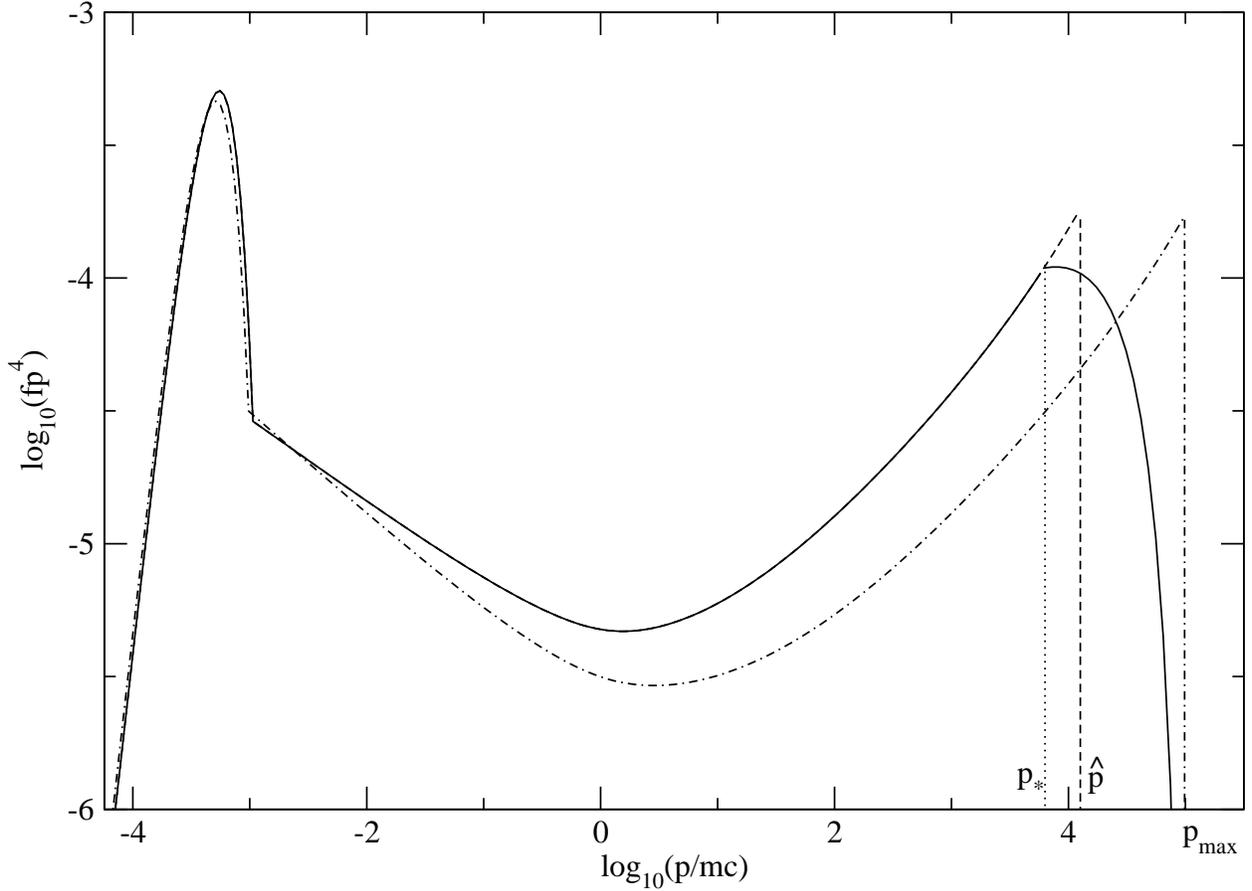}

\caption{Particle spectra at  a strong shock obtained from analytic solution
\citep{m97a, mdru01} for \protect\( M=150\protect \) (as in Figure~\ref{fig:bif})
and the injection rate \protect\( \nu \approx 0.1\protect \). The
dash-dotted line shows the solution with the abrupt momentum cut-off
at \protect\( p=p_{max}=10^{5}\protect \). The spectrum drawn with
the full line demonstrates the effect of wave compression
calculated using formulae (\ref{G0:sol}) and (\ref{nu:gen}).
The spectrum that would be obtained using the same technique as for
the dash-dotted case but with the maximum momentum at \protect\( p=\hat{p}\protect \)
(see text) is shown by the dashed line. \label{fig:match} }
\end{figure}

There are at least two reasons to believe that the standard acceleration
theory may have estimated the maximum particle energy or the form
of the spectrum below it incorrectly. The first reason is simply a
possible conflict with the observations of TeV-emission from SNRs
as we discussed in the Introduction section. The second reason is
a theoretical one, that arises naturally from considering the nonlinear
response of the shock structure to the acceleration which is exemplified
in Figure~\ref{fig:bif}. According to this picture, the response is
so strong that it is unlikely that the acceleration can proceed at
the same rate with no change in physics after such a dramatic shock
restructuring (pre-compression \( R \) may rise by 1-2 orders of
magnitudes depending on the Mach number). Time dependent numerical
simulations (\eg \citealt{kjg02}) show that the modifications occur
very quickly, and compression is increased substantially even before
\( p_{max}\sim 1 \) (note that this would be consistent with the
bifurcation diagram in Figure~\ref{fig:bif} for initial \( \nu \sim (c/u_{1})n_{CR}/n_{1}\ga 0.1 \),
where \( n_{CR}/n_{1} \) is the ratio of CR number density at the
shock to that of the background plasma far upstream). The shock modification,
in turn, must follow rather abruptly after the maximum momentum has
passed through the critical value. It was argued recently (\citealt{mdv00})
that this should drive crucial acceleration parameters such as the
maximum momentum and injection rate back to their critical values
which must limit shock modification and settle it at some marginal
level, the so called self-organized critical (SOC) level (see also
\citealt{tom00, mdru01, mdFer01} for more discussions of the critical
interrelation between the injection, maximum energy and shock structure).
Mathematically, the SOC state is characterized by the requirement
of merging of the two critical points on the bifurcation diagram in
Figure~\ref{fig:bif} into one inflection point on the \( \nu (R) \)
graph. Perhaps the most appealing aspect of this approach is its ability
to predict the values of all three order parameters (injection rate,
maximum momentum and compression ratio) given the only control parameter
(the Mach number) just from our knowledge of the nonlinear response
\( R(\nu ,p_{max}) \) shown in Figure~\ref{fig:bif}, and no further
calculations. 

However, the required backreaction mechanisms on the injection and
maximum momentum need to be demonstrated to operate. We have already
discussed at a qualitative level how the injection rate is reduced
by shock modification. The subject of this paper has been the reduction
of particle momenta related to the formation of a spectral break at
\( p=p_{*} \), as a result of wave compression in a modified shock
precursor. The position of the spectral break is universally related
to the degree of system nonlinearity \( R \), since \( p_{*}=p_{max}/R \).
Hence, the problem seems to be converted to the study of nonlinear
properties of the acceleration that are formally known from the analytic
solution shown in Figure~\ref{fig:bif}. However, the injection rate
\( \nu  \) that is now required for accurate determination of the
spectral break \( p_{*} \) through \( R \), may currently be obtained
only from the SOC ansatz. It should be also mentioned that strong
reduction of \( p_{*} \) is obviously not to be expected in oblique
shocks, where the resonance relation \( kp\propto B \) is approximately
preserved due to the compression of \( B \) simultaneously with \( k \). 

An equally important problem is that strong losses of particles between
\( p_{*} \) and \( p_{max} \) must slow down the growth of \( p_{max}(t) \)
due to the reduction of resonant waves. As we argued in section~\ref{sec:signif},
this may result in an order of magnitude slower acceleration than
one would expect from the standard Bohm diffusion paradigm. Consequently
the dynamically and observationally significant spectral break \( p_{*} \)
may be at least two orders of magnitude below the maximum momentum
\( p_{max} \) (again, depending on \( M \)) that could be reached
in the unimpeded acceleration which is normally implied in estimates
of maximum energy achievable in SNRs over their active life time. 

In addition to the above mentioned uncertainty in \( p_{max}(t) \),
its relation to the position of the spectral break \( p_{*}=p_{max}/R \)
also needs further clarification. Indeed, since \( R \) depends on
a dynamical cut-off \( \hat{p} \) which, in general, is linked to
\( p_{*} \) and \( p_{max} \), the latter relation is still implicit.
It can be easily resolved, however, in a supercritical regime (the
saturated part of the \( R(\nu ) \) dependence in Figure~\ref{fig:bif},
see also \citealt{mdru01} for details), which requires\footnote{%
This is strictly valid for \( \kappa (p)\propto p \).
} \( \nu \hat{p}/p_{inj}\gg M^{3/4} \). One simply has then \( R\approx M^{3/4} \).
As it was argued, however, the injection is unlikely to be high enough
to reach this regime. An additional argument against it is that the
spectral break becomes unrealistically small in the \( M\rightarrow \infty  \)
limit, since \( p_{*}=p_{max}/M^{3/4} \). In the opposite case \( \nu \hat{p}/p_{inj}\ll M^{3/4} \),
the compression rate saturates at \( R\approx \nu \hat{p}/p_{inj} \).
Note that the injection rate must be still above critical, otherwise
\( R\approx 1 \). Now we need to specify \( \hat{p} \). The simple
approximation used in the previous section yielded \( \hat{p}\approx 0.1p_{max} \),
so that \( p_{*}\approx 10p_{inj}/\nu  \) (independent of \( p_{max} \))
which may be regarded as a lower bound on \( p_{*} \). Indeed, the
above relation between \( \hat{p} \) and \( p_{*} \) may be applied
only to the outermost part of the shock transition (see Secs. \ref{sec: ext},\ref{sec:connect}).
Downstream, the spectrum cuts off very sharply immediately beyond
\( p_{*} \), section~\ref{sec: int}. Therefore, the dynamical cut-off
\( \hat{p}\approx p_{*} \) and we obtain the following upper bound
on \( p_{*} \), \( p_{*}\approx \sqrt{p_{inj}p_{max}(t)/\nu } \). 

It should be clear that unless \( \nu  \) is dramatically reduced
as a result of shock modification, even this upper bound places \( p_{*} \)
way below \( p_{max} \). This may be the reason for non-detection
of protons at TeV energies in SNRs. Finally, this does not contradict
to the detection of 10-100 TeV electrons in \eg SNR 1006 since they
may be accelerated by other mechanisms (\eg \citealt{pap81, gal84, bykuv99, lam01})
or may have higher radiation efficiency.

\acknowledgements{}

This study is supported through UCSD by the US Department of Energy,
Grant No. FG03-88ER53275. At the University of Minnesota this work
is supported by NASA through grant NAG5-8428, and by the University
of Minnesota Supercomputing Institute.

\clearpage

\clearpage

\clearpage

\begin{figure}
\plotone{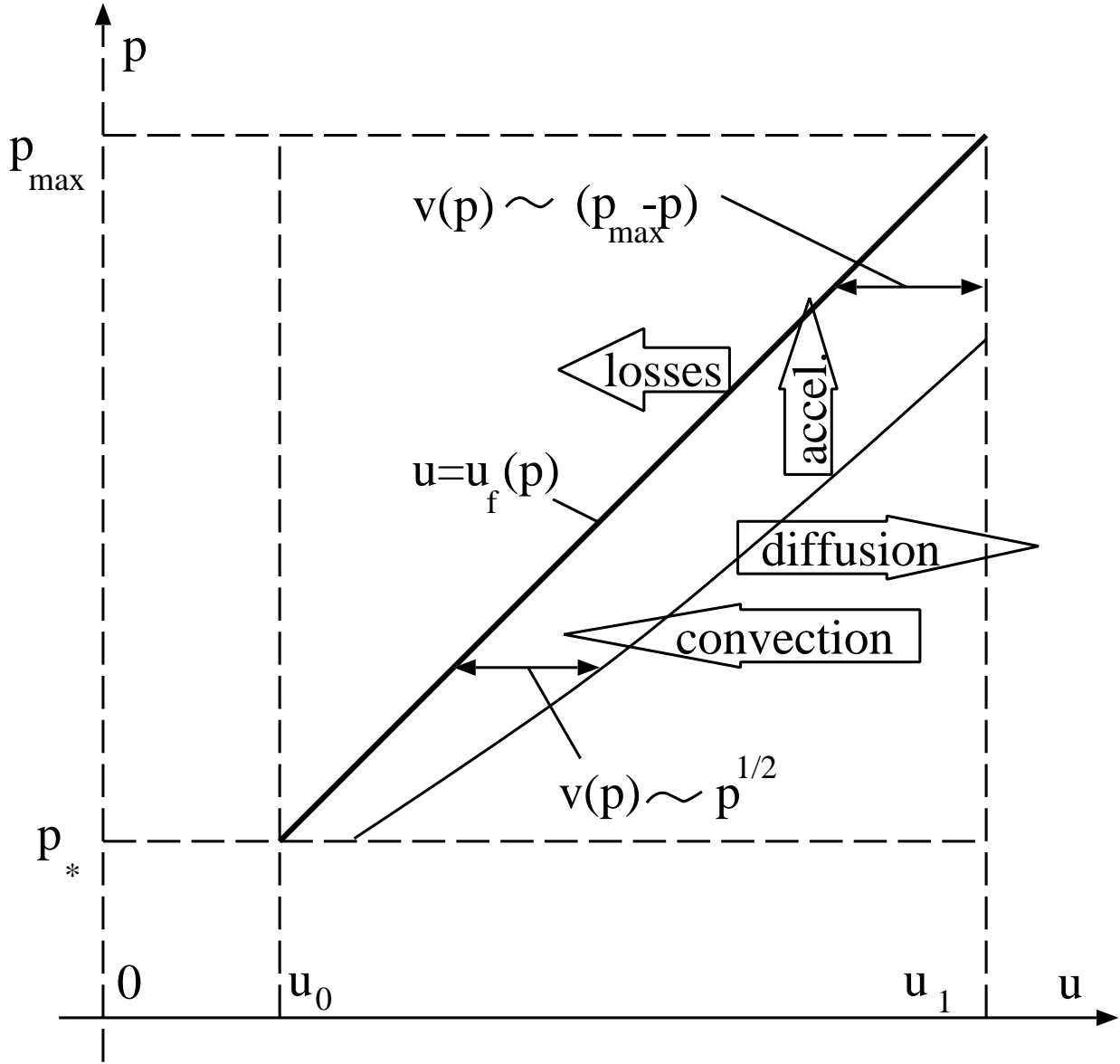}

\caption{The phase plane of accelerated particles in the flow velocity-particle
momentum coordinates. The particles are localized between the heavy
line (above which there are no resonant waves to confine them) and
the light line where the particle density decays exponentially towards
higher \protect\( u\protect \) (see text). The relevant transport
processes are indicated by arrows.\label{fig:fluxes}}
\end{figure}

\clearpage

\end{document}